\documentclass[prd,onecolumn,amsmath,amssymb,floatfix,nofootinbib]{revtex4}
\pdfoutput=1
\usepackage{graphicx}

\usepackage{amssymb}
\usepackage{amsmath}
\usepackage{feynmp}
\usepackage{slashed}

\usepackage{graphics}
\usepackage{dcolumn}
\usepackage{color}
\usepackage{rotate}
\usepackage{fancyhdr}
\usepackage{indentfirst}

\usepackage{umoline}
\usepackage{ulem}

\def\be{\begin{eqnarray}}
\def\ee{\end{eqnarray}}
\def\no{\nonumber}

\definecolor{darkred}{rgb}{.743,0,0}

\begin{document}
\title{On a possible large width 750~GeV diphoton resonance at ATLAS and CMS}

\author{Daniel Aloni$^{1}$, Kfir Blum$^{1}$, Avital Dery$^{1}$, Aielet Efrati$^{1}$, Yosef Nir$^{1}$}
\affiliation{$^1$Department of Particle Physics and Astrophysics, Weizmann Institute of Science, Rehovot, Israel 761000}
\email{daniel.aloni, kfir.blum, avital.dery, aielet.efrati, yosef.nir@weizmann.ac.il}
\date{\today}

\begin{abstract}
The ATLAS and CMS experiments at the LHC have reported an excess of diphoton events with invariant mass around 750~GeV, with local significance of about $3.6~\sigma$ and $2.6~\sigma$, respectively. We entertain the possibility that this excess is due to new physics, in which case the data suggest a new particle with 13~TeV LHC production cross section times diphoton branching ratio of about 5~fb. Interestingly, ATLAS reports a mild preference for a sizeable width for the signal of about 45~GeV; this result appears consistent with CMS, and is further supported by improving the compatibility of the 8~TeV and 13~TeV analyses. We focus on the possibility that the new state is a scalar. First, we show that, in addition to the new state that is needed directly to produce the diphoton bump, yet more new particles beyond the Standard Model are needed to induce diphoton decay rate of the right size. Second, we note that if the excess is attributed to the Breit-Wigner peak of a single new state, then the signal strength and width -- taken together -- suggest a total LHC production cross section of order $10^5$~fb. Restricting to perturbative models without ad-hoc introduction of many new states or exotic charges, we reach the following conclusions: (i) Gluon-fusion cannot explain the required large production cross section. (ii) Tree level production from initial state quarks cannot explain the required branching ratio to two photons. (iii) Tree level production is constrained by flavor data as well as LHC Run-I and Tevatron dijet analyses. Insisting on a large width we are led to suggest that more than one scalar states, nearly degenerate in mass, could conspire to produce an observed wide bump.
\end{abstract}

\maketitle

\tableofcontents

\section{Introduction}

The Large Hadron Collider (LHC) experiments, ATLAS and CMS, reported an excess of diphoton events in their Run-II 13~TeV analyses, using 3.2fb$^{-1}$ and 2.6fb$^{-1}$ of data, respectively~\cite{ATLAS-CONF-2015-081,CMS-PAS-EXO-15-004}. The local and global significance of the excess are estimated by $\sim3.6\sigma~(2.6\sigma)$ and $2\sigma~(1.4\sigma)$, respectively, at ATLAS (CMS). The occurrence of an excess in both ATLAS and CMS, compatible with Run-I results, makes an exciting case for new physics. Here we assume that this is indeed the case.
The reported excess is centered around $M_S\approx750$~GeV, and requires a 13~TeV cross section around $\sigma(pp\to S\to\gamma\gamma)\approx2-8$~fb, where $S$ is either a scalar or a spin-two particle produced in s-channel; in what follows we focus on the scalar possibility.

Interestingly, ATLAS reports a pull in the 13~TeV data towards a sizeable signal width\footnote{See Sec.~10 in ~\cite{ATLAS-CONF-2015-081}. The effect is reported as a $1.5~\sigma$ systematic pull in the nuisance parameter associated with the photon energy resolution uncertainty.}, with best fit $\Gamma_S=45$~GeV$\approx0.06M_S$.
A finite width is also somewhat preferred, by about 0.8$\sigma$ compared to the narrow width alternative, for compatibility with Run-I results.
In terms of the theoretical interpretation, there is crucial distinction between the large and narrow width possibilities, suggesting very different model-building avenues. It is therefore interesting to clarify the theoretical perspective on this issue, even while the observational evidence is still inconclusive. This is our main objective in the current paper. We combine complimentary constraints from collider and precision flavor physics, showing that a large intrinsic width for the new particle $S$ would pose a generic theoretical challenge for perturbative model interpretations. We show how relaxing the large width assumption alleviates these difficulties, making the reported signal width a key experimental observable for theory interpretation. In addition to the issue of signal width, we show that new states beyond the SM, in addition to the $S$ particle itself, are required to explain the coupling of $S$ to photons.

Our reasoning is as follows. The partial production cross section in the diphoton channel is related to the total cross section by\footnote{The expression we use is applicable in the narrow width approximation, that is valid to about ten percent accuracy even if we consider total width $\Gamma_S$ larger by a factor of two or so than the ATLAS best fit value. Thus even though we refer to $\Gamma_S\approx0.06M_S$ as ``large width" in the text, what we mean by large width -- as will become clear in discussing the physics implications -- is essentially $\Gamma_S\gtrsim0.01M_S$ or so.} $\sigma(pp\to S)\approx\frac{\sigma(pp\to S\to\gamma\gamma)}{BR(S\to\gamma\gamma)}$. This gives the constraint
\be\label{eq:sigtree}
\left[\frac{\sigma(pp\to S)}{1~{\rm pb}}\right]\times\left[\frac{\Gamma(S\to\gamma\gamma)/M_S}{\left(\alpha/4\pi\right)^2}\right]\approx770~\left[\frac{\sigma(pp\to S\to\gamma\gamma)}{5~{\rm fb}}\right]\times\left[\frac{\Gamma_S/M_S}{0.06}\right],\ee
at the 13~TeV LHC. On the right, we put the observational data scaled to the best fit values. On the left, we put the implied theory requirement. Here we scale the total production cross section of $S$ by 1~pb, comparable to the 0.74~pb NNLO ggF production cross section~\cite{Heinemeyer:2013tqa} for a heavy SM-like Higgs boson at 750~GeV.
Concerning the diphoton decay width, we note that perturbative models, without baroque model building, predict\footnote{We will get back to this point with explicit examples later on, e.g. in App.~\ref{app:2hdm}. The more explicit dependence on parameters is $\Gamma(S\to\gamma\gamma)\sim \left(\frac{\alpha}{4\pi}\right)^2\frac{M_S^3}{M^2}N^2Q^4$, where $M$ is the mass of some other new particle running in the $S\gamma\gamma$ loop amplitude, with EM charge $Q$ and multiplicity $N$. We are assuming mundane $Q,N=\mathcal{O}(1)$, as well as $M\sim M_S$, since both $M\ll M_S$ and $M\gg M_S$ would suppress the amplitude, requiring an even larger total production cross section in Eq.~(\ref{eq:sigtree}).} $\Gamma(S\to\gamma\gamma)\lesssim \left(\frac{\alpha}{4\pi}\right)^2M_S$. Note that in Eq.~(\ref{eq:sigtree}) we do not assume s-channel exchange, or indeed any particular partonic topology, for the production process of $S$ at the LHC.

We find that a width of $\Gamma_S\sim0.06M_S$ would require either very large total production cross section, or very large partial decay width to photons, or some large combination of both. This model-independent constraint summarizes the basic model-building challenge in explaining a large intrinsic total width for $S$.

In Sec.~\ref{sec:npdec}, before going into the signal width issue, we show that existing Run-I constraints on $t\bar t$ and $WW$ production at the 8~TeV LHC imply that new states, in addition to $S$ itself, must be introduced to facilitate a large enough $S\gamma\gamma$ coupling.
In Sec.~\ref{sec:1} we show that the cross section given by Eq.~(\ref{eq:sigtree}) is difficult to achieve with a gluon fusion (ggF) loop, unless the true width is smaller than the $\Gamma_S=45$~GeV best-fit ATLAS value by about two orders of magnitude. Forced to consider tree level production of a 750~GeV new state, we are also led to a tight spot in model building with various constraints including flavor and compatibility with Run-I.
We summarise our results and conclude in Sec.~\ref{sec:conc}. In App.~\ref{app:a} we recap some loop functions used in the analysis. App.~\ref{app:2hdm} illustrates the theory difficulty in obtaining $\frac{\Gamma(S\to\gamma\gamma)/M_S}{\left(\alpha/4\pi\right)^2}>1$ within the specific example of two Higgs doublet models.

\section{New physics states in addition to $S$ are needed {\it at least} for the diphoton decay amplitude}\label{sec:npdec}

New particles beyond the SM, in addition to the scalar $S$, are needed to explain the effective coupling of $S$ to photons. This is exciting news: it means that $S$ itself is probably the tip of the iceberg and more new physics states, charged under EM, await discovery. To derive this result, note that the only SM particles that could be relevant to the $S\gamma\gamma$ loop amplitude are the top quark and the $W^\pm$ boson, as all other charged states in the SM have small masses $m\ll 0.1M_S$, resulting with strong suppression from the loop amplitude. Since $M_S\gg 2m_t,\;2m_W$, utilising $t$ and/or $W^\pm$ in the $S\gamma\gamma$ loop amplitude implies that the tree level decays $S\to t\bar t$ and/or $S\to W^+W^-$ are open, and are related to the $S\gamma\gamma$ amplitude by one and the same underlying coupling. This point brings about constraints from Run-I $W$ and top pair production, that exclude the $S\gamma\gamma$ amplitude from being explained entirely by SM contributions.
\begin{figure}[htbp]
\includegraphics[width=0.35\linewidth]{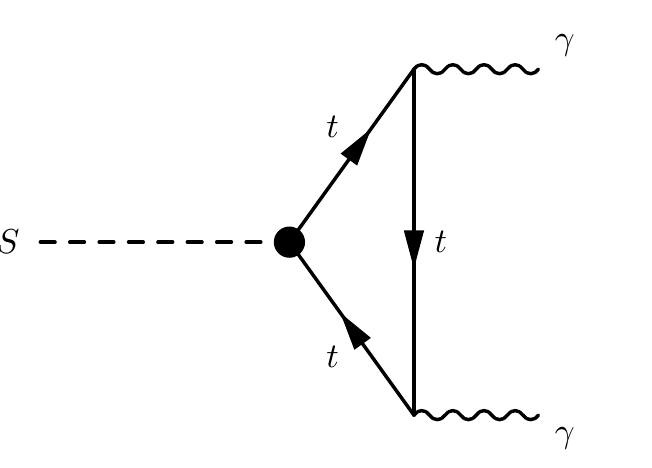}
\includegraphics[width=0.35\linewidth]{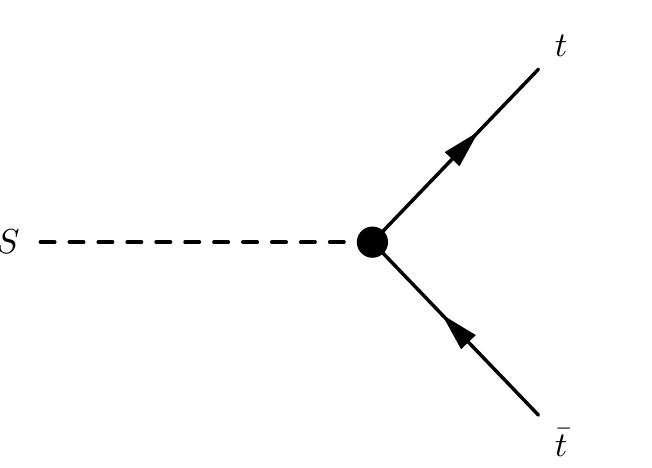}
\caption{Feynmann graphs illustrating the connection between the loop contribution of SM top quarks in the $S\to\gamma\gamma$ loop amplitude, and the tree-level $S\to t\bar t$ decay proceeding via the same $Stt^c$ coupling.}
\label{fig:feyn}
\end{figure}

We bound the top quark contribution to $S\to\gamma\gamma$ using the Run-I 8~TeV ATLAS and CMS $t\bar t$ analyses~\cite{Aad:2015fna,Khachatryan:2015sma}. For our purpose in this section, these analyses can be summarized by the approximate bound $\sigma_8(pp\to S\to t\bar t)<1$~pb at 95\%CL. (Here and elsewhere, a subscript on a cross section means that the cross section applies to the LHC with the stated center of mass energy. Cross section without subscript should be understood to apply to the 13~TeV LHC.)

To induce the diphoton effect we need a coupling between $S$ and the top,
\be\label{eq:stt}\mathcal{L}=-yStt^c+{\rm h.c.}\ee
We can take $S$ to be an SU(2) doublet, in which case $y$ is a dimensionless Yukawa coupling, but we could also allow $S$ to be {\it e.g.} an SM-singlet, in which case the interaction above would have dimension five at least, namely $y=\tilde yv/\Lambda$, where $v$ is the SM Higgs VEV, $\Lambda$ some effective field theory (EFT) cut-off scale, and $\tilde y$ a dimensionless coupling. Our results apply equally to these possibilities.

Eq.~(\ref{eq:stt}) induces both an effective coupling of $S$ to the photons from the top loop, and tree-level decay of $S$ to $t\bar t$. This is illustrated by the Feynmann graphs in Fig.~\ref{fig:feyn}. We find,
\be
\Gamma(S\to\gamma\gamma)&=&\frac{4\alpha^2}{81\pi^3}|y|^2\,\tau|A_t(\tau)|^2\,M_S,\\
\Gamma(S\to t\bar t)&=&\frac{3}{8\pi}|y|^2\,M_S\left(1-\frac{4m_t^2}{M_S^2}\right)^{\frac{3}{2}}.
\ee
We recall the loop amplitude $\mathcal{A}_\psi(\tau)$, $\tau\equiv \left(\frac{M_S}{2m_\psi}\right)^2$, in App.~\ref{app:a}. The phase space factor in the decay $S\to t\bar t$ is $\left(1-4m_t^2/M_S^2\right)^{\frac{3}{2}}\sim0.7$.
Note that the total production cross section for $S$ at the 8~TeV LHC is related to that at 13~TeV via $\sigma_{13}(pp\to S)/\sigma_{8}(pp\to S)=\mathcal{L}_{13}/\mathcal{L}_{8}$, where $\mathcal{L}_{13}/\mathcal{L}_{8}$ is the corresponding ratio of parton luminosities relevant for the production of $S$. Combining these results we find
\be\label{eq:ttlim}\sigma_{13}(pp\to S\to\gamma\gamma)&=&\left(\frac{\mathcal{L}_{13}}{\mathcal{L}_{8}}\right)\sigma_8(pp\to S\to t\bar t)\frac{\Gamma(S\to\gamma\gamma)}{\Gamma(S\to t\bar t)}\no\\
&\lesssim&3.9\times10^{-2}\left(\frac{\mathcal{L}_{13}/\mathcal{L}_{8}}{5}\right)~{\rm fb},\ee
where the inequality comes from imposing our rough 1~pb $t\bar t$ limit based on~\cite{Aad:2015fna,Khachatryan:2015sma}.
Note that the coupling $y$ cancels in the partial decay width ratio $\Gamma(S\to\gamma\gamma)/\Gamma(S\to t\bar t)$.
We scaled this ratio of parton luminosities by a number close to the ggF value of 4.7, that would apply for s-channel production of $S$, noting that this ratio for quark-gluon fusion and qq is smaller than that for ggF.

From Eq.~(\ref{eq:ttlim}) we conclude that, at least considering direct s-channel production of $S$, the decay to two photons cannot be mediated by a top loop alone, as that would lead to violation of the 8~TeV $t\bar t$ limits of Refs.~\cite{Aad:2015fna,Khachatryan:2015sma} by about two orders of magnitude.

\begin{figure}[htbp]
\includegraphics[width=0.45\linewidth]{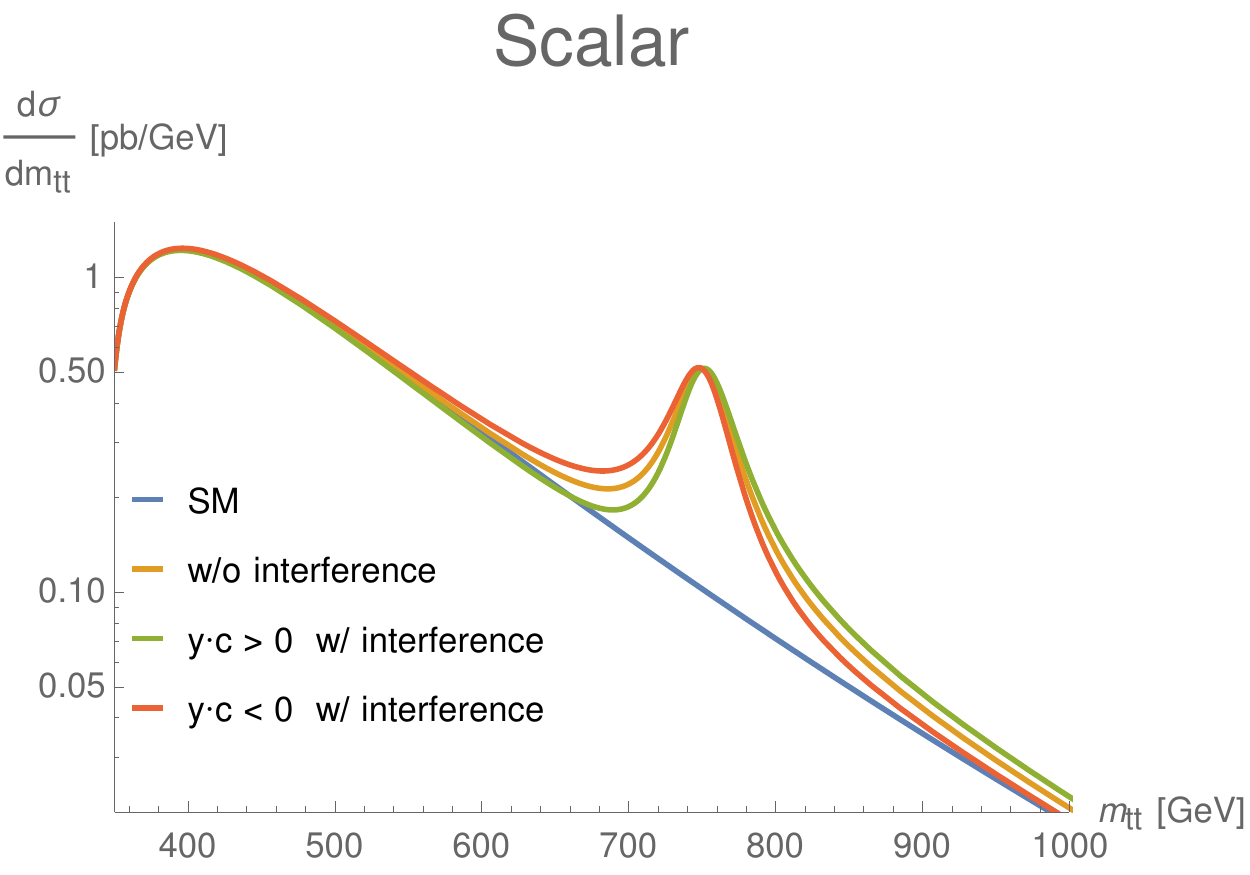}
\includegraphics[width=0.45\linewidth]{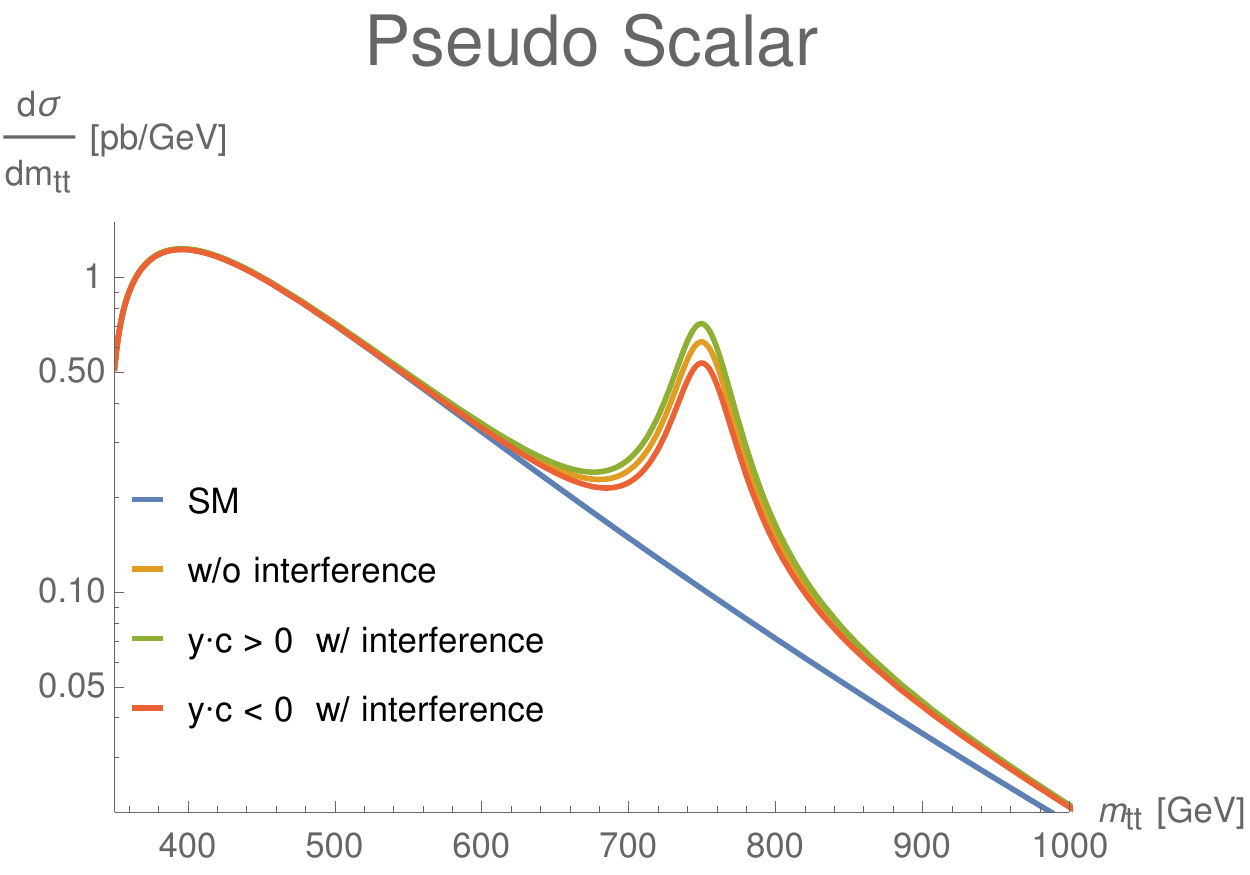}
\caption{The $t\bar t$ differential cross section for the SM and adding a scalar (Left) and pseudoscalar (Right) $S$, inducing the diphoton excess via ggF production (parameterized by an effective $SGG$ operator) and top loop-induced diphoton decay arising from the interaction in Eq.~(\ref{eq:stt}). The parameters are chosen to give $\sigma_{13}(pp\to S\to\gamma\gamma)=2$~fb, at the lower end required to address the diphoton excess.}
\label{fig:int}
\end{figure}

Interference effects between the SM and new physics contributions to $t\bar t$ production could be significant for new physics contributions to the differential cross section of the same order of magnitude as the SM one. Eq.~(\ref{eq:ttlim}), however, relates a top-mediated $S\to\gamma\gamma$ decay to factor of 10-100 excess in $t\bar t$ cross section. This result is therefore insensitive to the interference effects. To clarify this point, in Fig.~\ref{fig:int} we show the 8~TeV LHC $t\bar t$ differential cross section, calculated for the SM alone and for (left) an additional scalar or (right) pseudoscalar particle $S$ with parameters chosen to account for the diphoton excess. We use an effective $S$-gluon coupling $cSG_{\mu\nu}G^{\mu\nu}$ ($cSG_{\mu\nu}\tilde G^{\mu\nu}$ for pseudo-scalar $S$) to induce ggF production of $S$ (interference effects are maximized if the production mechanism is ggF), and the interaction of Eq.~(\ref{eq:stt}) to induce diphoton decay (with real $y$ for scalar or $iy$ for pseudo-scalar). Our matrix element is computed at LO with FeynCalc~\cite{Mertig:1990an,Shtabovenko:2016sxi} and convolved with MSTW2008 NNLO pdf~\cite{Martin:2009iq}. 
We use a flat K-factor of $1.86$ to normalize the SM total cross section to NNLO calculations \cite{Czakon:2013goa}.

\begin{figure}[htbp]
\includegraphics[width=0.465\linewidth]{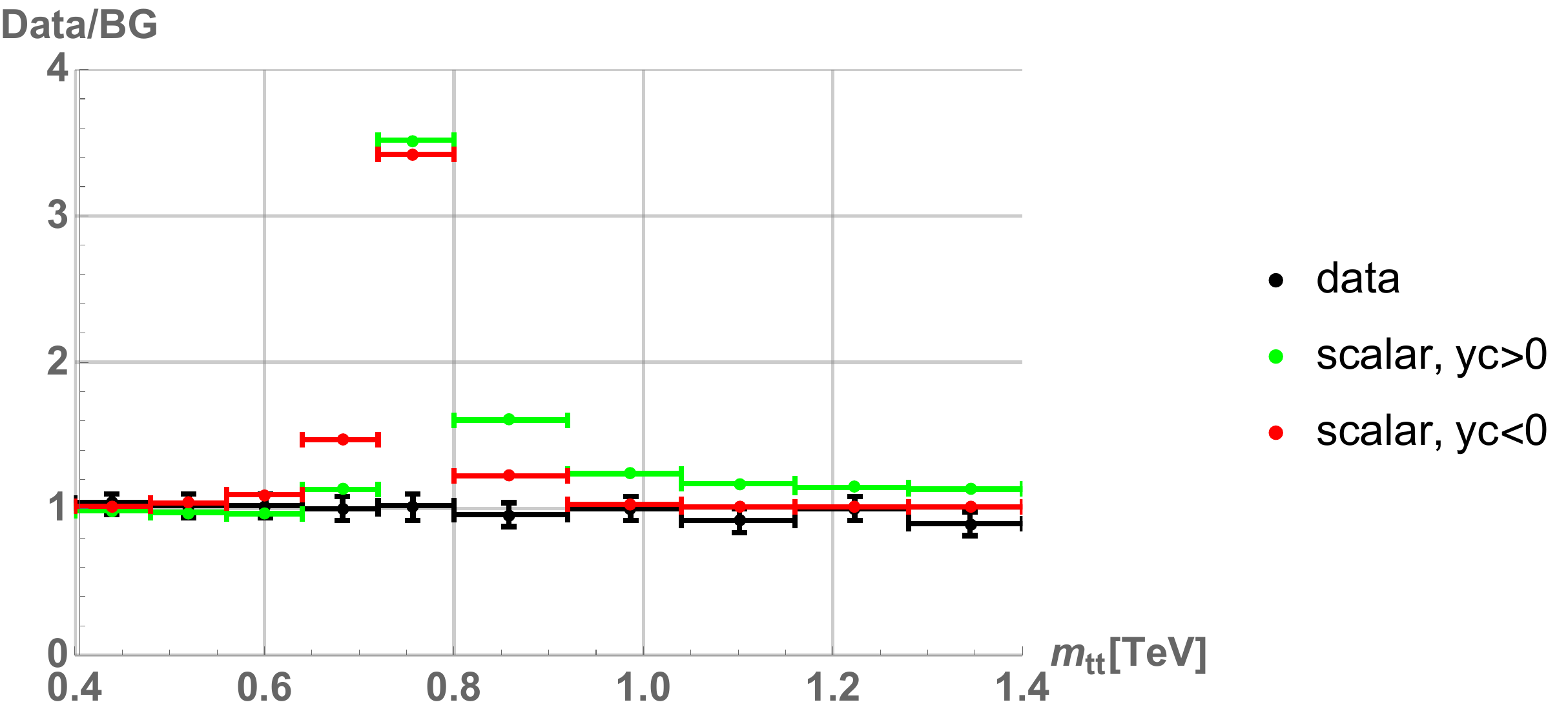}
\includegraphics[width=0.495\linewidth]{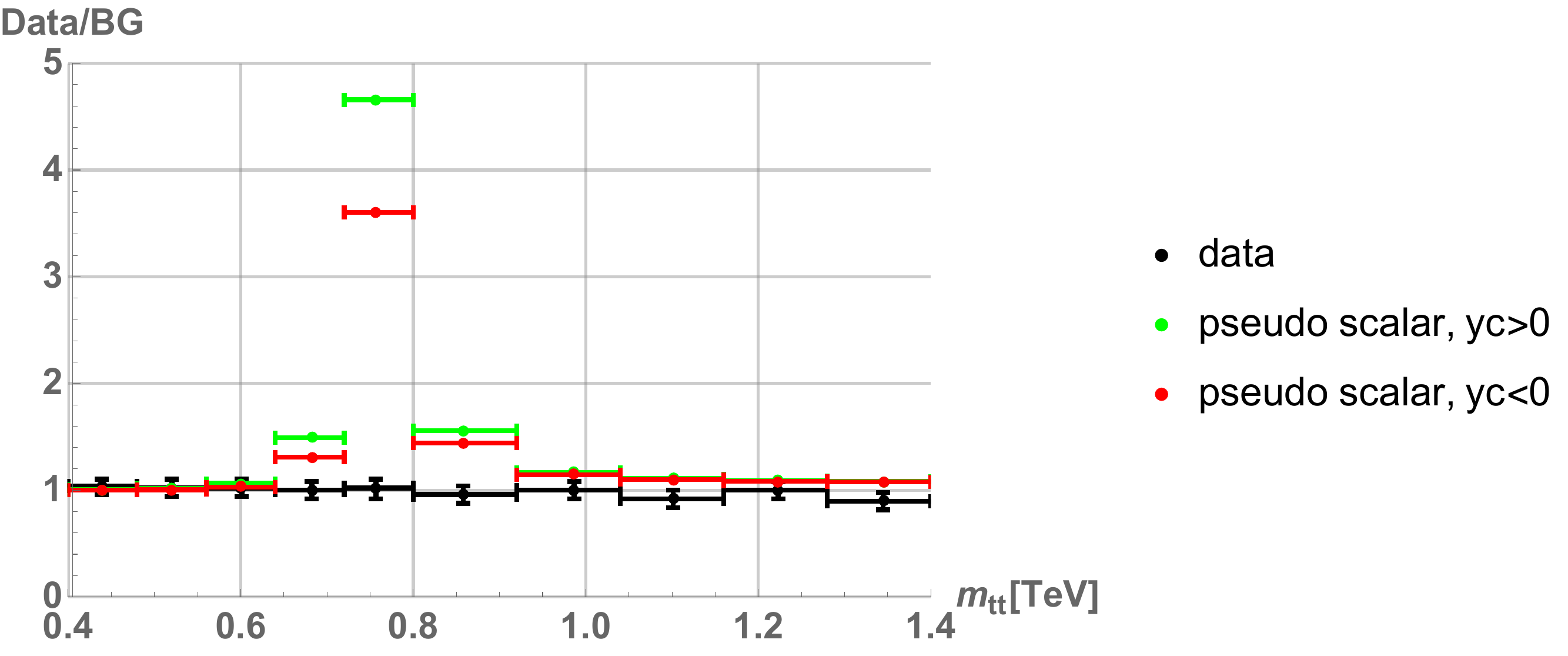}
\caption{Binned $t\bar t$ distribution, showing the ratio in each bin to the SM prediction. Black data markers are from the ATLAS analysis of~\cite{Aad:2015fna}. Red and green markers show the result for the scalar (Left) and pseudoscalar (Right) models as in Fig.~\ref{fig:int}, adjusting parameters to obtain $\sigma_{13}(pp\to S\to\gamma\gamma)=2$~fb, at the lowest end of the cross section required for the diphoton excess. The red and green colors correspond to opposite sign assumption for the product of couplings controlling the interference between the SM and new physics amplitudes. These models are ruled out by the data, as expected from Eq.~(\ref{eq:ttlim}).}
\label{fig:int2}
\end{figure}

The set of three bumps corresponds to adjusting the parameters to $\sigma_{13}(pp\to S\to\gamma\gamma)=2$~fb, at the lowest end of the cross section required for the diphoton excess. The difference between these three curves corresponds to different choices for the sign of the product of couplings $yc$, with the middle curve corresponding to artificially removing interference by hand. Clearly interference, being a ten percent effect in the relevant parameter space, is unimportant to our basic conclusion that top-induced diphoton decay would imply a major discrepancy in the $m_{t\bar t}$ distribution. In Fig.~\ref{fig:int2} we further compare the $t\bar t$ distributions to data, using for concreteness the measurements in Ref.~\cite{Aad:2015fna} presenting the binned $m_{t\bar t}$ distribution divided by the expected SM prediction in each bin. The plot assumes $\sigma_{13}(pp\to S\to\gamma\gamma)=2$~fb, again at the lowest end to address the diphoton excess. Top-mediated diphoton decay is unambiguously excluded by data.

Finally, while $\mathcal{L}_{13}/\mathcal{L}_{8}\lesssim5$ would apply for s-channel direct production of $S$, we can contemplate the possibility that $S$ is produced in association with some other massive new physics particle with mass $M>M_S$, that then decays into $S$. In this case, the ratio $\mathcal{L}_{13}/\mathcal{L}_{8}$, that we scale in Eq.~(\ref{eq:ttlim}) by a value relevant for 750~GeV center of mass energy, could be larger than 5, potentially relaxing the constraint. Assuming for definiteness that the parent state is pair-produced\footnote{The argument follows similarly if the heavy parent state is singly produced, though one would then need to require $M\gtrsim4$~TeV to obtain the required increase in parton luminosity ratio from 8 to 13~TeV. We do not discuss here additional collider signatures of such scenario, including the expectation that $S$ would be born significantly boosted in the partonic center of mass frame, or the possibility of  missing transverse energy or additional hadronic activity -- none of which where reported in~ Ref.\cite{ATLAS-CONF-2015-081} -- following the heavy state decay.}, we note that to obtain $\sigma_{13}(pp\to S\to\gamma\gamma)\gtrsim2$~fb, namely above the lower range required for the signal, one would need $\mathcal{L}_{13}(2M)/\mathcal{L}_{8}(2M)\gtrsim400$. This translates to $M\gtrsim2$~TeV. At such large $M$, however, we expect that the LHC production cross section for the heavy states $M$ is already quite small even at tree level QCD. For example, for this mass, gluino pair production has $\sigma(pp\to\tilde g\tilde g)$ in the ballpark of 1~fb. This small cross section leaves no room for the additional parametric $\sim(\alpha/4\pi)^2$ suppression of $BR(S\to\gamma\gamma)$. We conclude that it is highly unlikely for the SM top quark to be the sole mediator of $S\to\gamma\gamma$ decay.

Analogously to the top quark discussion above, the $W^\pm$ contribution to $S\to\gamma\gamma$ can be bounded using the Run-I  8~TeV ATLAS $WW$ analysis~\cite{Aad:2015agg}, from which we have $\sigma_8(pp\to S\to WW)<0.1$~pb at 95\%CL. Adopting the simple model $\mathcal{L}=\kappa SW^+_\mu W^{-\mu}$, we find\footnote{The SM Higgs would have, in this notation, $\kappa=gm_W$, with the electroweak coupling $g=0.65$; while for a heavy $S$ we expect a further suppression factor of $\sim v^2/M^2_S$ in $\kappa$. We note that the $W^\pm$ loop amplitude contributing to the effective $S\gamma\gamma$ decay rate is suppressed, relative to the corresponding $t\bar t$ contribution, by a factor $\sim0.1$, due to the smaller ratio $m_W^2/M_S^2$. Also note that, with our assumed coupling, the $S\to WW$ decay is dominated by $W$ longitudinal polarization.},
\be
\Gamma(S\to\gamma\gamma)&=&\frac{\alpha^2}{256\pi^3}\frac{|\kappa|^2}{m_W^2}\,\tau|A_W(\tau)|^2\,M_S,\\
\Gamma(S\to WW)&\approx&\frac{1}{64\pi}\frac{|\kappa|^2}{m_W^2}\,\frac{M_S^3}{m_W^2}\sqrt{1-\frac{4m_W^2}{M_S^2}}\left(1-\frac{4m_W^2}{M_S^2}+\frac{12m_W^4}{M_S^4}\right).
\ee
We recall the loop amplitude in App.~\ref{app:a}. Combining these results we find
\be\label{eq:wwlim}\sigma_{13}(pp\to S\to\gamma\gamma)&=&\left(\frac{\mathcal{L}_{13}}{\mathcal{L}_{8}}\right)\sigma_8(pp\to S\to WW)\frac{\Gamma(S\to\gamma\gamma)}{\Gamma(S\to WW)}\no\\
&\lesssim&1.2\times10^{-3}\left(\frac{\mathcal{L}_{13}/\mathcal{L}_{8}}{5}\right)~{\rm fb}.\ee
Repeating the arguments of the $S\to t\bar t$ constraints we see that the $W^\pm$ loop cannot induce a large enough $S\gamma\gamma$ amplitude, falling short by about three orders of magnitude given current bounds from direct $pp\to WW$ measurements. Thus, new particles in addition to $S$ need to be added in consistent new physics explanations of the ATLAS and CMS diphoton excess.

\section{The model-building challenge of a large width}\label{sec:1}

A large decay width, as possibly hinted to by ATLAS data, would be difficult to account for with ggF loop production. In Sec.~\ref{ssec:loopwide} we substantiate this point with a toy model example, illustrating the generic problem. We then turn in Sec.~\ref{ssec:treewide} to examine the possibility of tree-level production of the 750~GeV state.

\subsection{Large total width poses a challenge for ggF loop production of $S$}\label{ssec:loopwide}
Consider a toy model with a real singlet scalar\footnote{We denote the quantum numbers of a field $\Phi$ w.r.t. the SM gauge groups by $\Phi(D_c,D_w)_Y$, where $D_c$ and $D_w$ are the representations under SU(3)$_C$ and SU(2)$_L$ and $Y$ is the hypercharge.} $s(1,1)_0$ together with color-triplet, EM-charge $Q$, vector-like fermions $\psi(3,1)_Q,\;\psi^c(\bar 3,1)_{-Q}$, such that
\be\mathcal{L}_{\rm toy}&=&-\frac{M_S^2}{2}s^2-\left\{\left(m_\psi+ys\right)\psi\psi^c+{\rm h.c.}\right\}.\ee
This model mimics the SM Higgs-top coupling. The effective Lagrangian, integrating out $\psi$ and $\psi^c$ at one loop, is (see e.g.~\cite{Kniehl:1995tn})
\be\mathcal{L}_{\rm toy,eff}&=&\frac{\alpha_s\,y}{12\pi\,m_\psi}N\mathcal{A}_\psi\,sG^a_{\mu\nu}G^{a\mu\nu}+\frac{\alpha\,y}{2\pi\,m_\psi}NQ^2\mathcal{A}_\psi\,sF^a_{\mu\nu}F^{a\mu\nu},\ee
where we allowed for $N$ identical copies of fermions.

The scalar decay width to two photons is
\be\frac{\Gamma(s\to\gamma\gamma)}{M_S}&=&N^2Q^4\left(\frac{\alpha}{4\pi}\right)^2\frac{4}{\pi}y^2\left|\mathcal{A}_\psi(\tau)\right|^2\tau<0.65y^2N^2Q^4\left(\frac{\alpha}{4\pi}\right)^2,\ee
using the fact that the factor $\left|\mathcal{A}_\psi(\tau)\right|^2\tau$, limiting to $m_\psi>500$~GeV (meaning $\tau<0.5$), is bounded by\footnote{The absolute maximum value of the factor $\left|\mathcal{A}_\psi(\tau)\right|^2\tau$ is $\approx6.85$, reached around $\tau\sim4$, or $M_S\sim4m_\psi$. But here $m_\psi\sim190$~GeV would be excluded by current collider limits.} $\left|\mathcal{A}_\psi(\tau)\right|^2\tau<0.5$.

The toy model at fixed $M_S=750$~GeV has two free parameters for ggF: $m_\psi/M_S$ and the coupling $y$. (The SM, of course, has $y$ fixed by $y_t=m_t/v$, where $m_t$ is the top quark mass.) The ggF production cross section for $s$ is identical to that of the SM Higgs $h$, up to the replacement $m_h\to M_S$ and $v\to m_\psi/y$, with $v=246$~GeV being the SM Higgs vacuum expectation value (VEV).
Comparing to the SM ggF cross section for a heavy Higgs, we therefore have
\be\label{eq:sigS}\sigma(gg\to s)&=&\sigma_{\rm SM}(gg\to h;\,m_h=M_S)\times\left|\frac{\mathcal{A}_\psi\left(\frac{M_S^2}{4m_\psi^2}\right)}{\mathcal{A}_\psi\left(\frac{M_S^2}{4m_t^2}\right)}\right|^2\frac{y^2v^2}{m_\psi^2}N^2.\ee

The LHC 13~TeV SM NNLO+NNLL result at $m_h\to750$~GeV is $\sigma_{SM}(gg\to h;\,m_h=750~{\rm GeV})\approx0.74$~pb (the LO result is smaller by a K-factor of 2.6). Limiting ourselves to perturbative $y<1$ and to $m_\psi\geq500$~GeV, as a conservative estimate to the lowest mass allowed by existing searches for new colored fermions, we find $\sigma(gg\to s)\leq0.2$~pb for $N=1$.
Comparing with Eq.~(\ref{eq:sigtree}), and allowing $m_\psi=500$~GeV, the requirement of the toy model to explain the ATLAS and CMS excess translates into
\be\label{eq:sigtoy}
y^4N^4Q^4\;\gtrsim\;6\times10^3~\left[\frac{\sigma(pp\to S\to\gamma\gamma)}{5~{\rm fb}}\right]\times\left[\frac{\Gamma_S/M_S}{0.06}\right],\ee
To accommodate Eq.~(\ref{eq:sigtoy}) at the best fit values of $\Gamma_S/M_S$ and $\sigma(pp\to S\to\gamma\gamma)$, without resorting to exotic charges $Q>1$, we would need, {\it e.g.}, 8 copies of the vector-like fermions, all with large $y\approx1$ Yukawa couplings and rather light mass $\sim$500~GeV. It is unlikely that such multitude of sub-TeV colored fermions would have evaded Run-I constraints.

Somewhat larger cross section could be obtained with more complex model-building, namely, adding more new colored states with large couplings to $s$ to increase the ggF production further, or adding several or exotically-charged uncolored EM-charged vector-like fermions to boost the diphoton decay.
However, the main message we see in this result is that Eq.~(\ref{eq:sigtree}), the generic impact of which is illustrated in our toy model constraint of Eq.~(\ref{eq:sigtoy}), highlights the width $\Gamma_S$ as a key observable in the theoretical interpretation of the excess: allowing $\Gamma_S$ to decrease by 2-3 orders of magnitude from its best fit value would make it rather easy to attribute the excess to ggF production in relatively minimal perturbative extensions of the SM.

On the other hand, without appealing to multi-component models and if we insist on believing that the large width $\Gamma_S\sim 0.06M_S$ is correct, then the ATLAS signal appears to call for tree level production of the new 750~GeV state. In the next section we study this direction, showing that it runs into difficulties of its own.

\subsection{Constraints on a large width with tree level production of $S$}\label{ssec:treewide}

To couple a scalar field at tree level to the initial state in pp collisions -- guessing in advance that sizable couplings are required, suggesting renormalizable models -- we would need a Higgs-like electroweak doublet with Yukawa couplings to the quarks, notably of the first generation.
We therefore consider a two-Higgs doublet model (2HDM): in addition to the usual light Higgs of the SM, we add the new field
$H(1,2)_{\frac{1}{2}}=\left(H^+,\frac{H_0+iA_0}{\sqrt{2}}\right)^T$.

We first consider the possibility of Yukawa coupling of $H^0$ to first generation up quarks.
Here, production at the LHC is dominated by
\be\label{eq:treeL}\mathcal{L}&=&(Hq)Yu^c+{\rm h.c.}\no\\
&=&H^+dV^TYu^c-\frac{H_0+iA_0}{\sqrt{2}}uYu^c+{\rm h.c.},\ee
written in the quark mass basis, with $V$ being the CKM matrix. We assume for simplicity that the couplings $Y$ are real.

Avoiding fine-tuning for the first generation quark masses requires
\be\langle H\rangle\ll v,\ee
where $v=246$~GeV and $\langle H\rangle$ are the SM-like Higgs and the new scalar vacuum expectation values, respectively.
While some mass mixing of $H$ with the SM Higgs $h$ is unavoidable, the details are model-dependent and we ignore this point here for simplicity.
We note that quartic terms of the form $\lambda H^2h^2$ could split the $H^0,\,A^0,\,H^\pm$ states at order $\sim100$~GeV (or perhaps more likely $\sim10$~GeV if we judge based on the very perturbative SM Higgs self-quartic $\lambda_h\approx0.1$). This is a parametrically small splitting compared to the scale $M_S$ and we ignore this complication in what follows in kinematics and loop functions, though we comment about other implications later on.

The decay width of $H^0$ to two quarks is
\be\Gamma(H^0\to u\bar u)&=&\frac{3|Y|^2}{16\pi}M_H=45\,|Y|^2\,\left(\frac{M_H}{750~{\rm GeV}}\right)~{\rm GeV},\ee
with a similar result for $A^0$.
We use MG5@NLO~\cite{Alwall:2014hca} to estimate the s-channel total production cross section of $H^0$ at the 13~TeV LHC, finding
\be\label{eq:1}\sigma(pp\to H^0)\approx125K|Y|^2~{\rm pb}\ee
at $M_H=750$~GeV, where $K$ is an order unity $K$-factor that we introduce here by hand. The result for $\sigma(pp\to A^0)$ is similar. We get a rough idea of the magnitude of the $K$ factor by computing the cross section $\sigma(pp\to H^0j)$, for which we get $\sigma(pp\to H^0j)\approx80$~pb. This suggests $K\sim1.5-2$. We use this estimate below, though our results are not very sensitive to the precise value of $K$ as long as it is $\mathcal{O}(1)$.

\begin{figure}[htbp]
\includegraphics[width=0.45\linewidth]{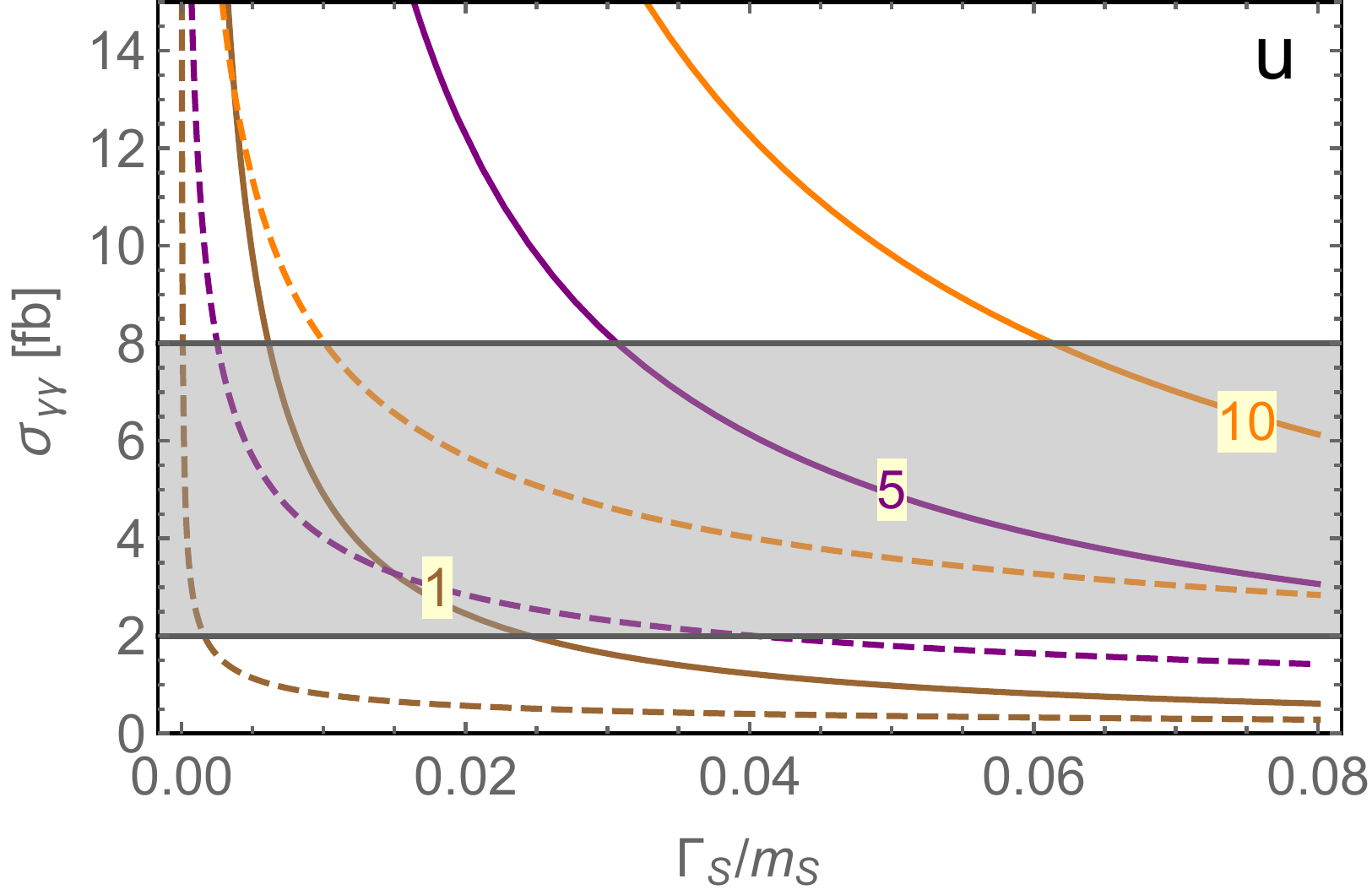}
\includegraphics[width=0.45\linewidth]{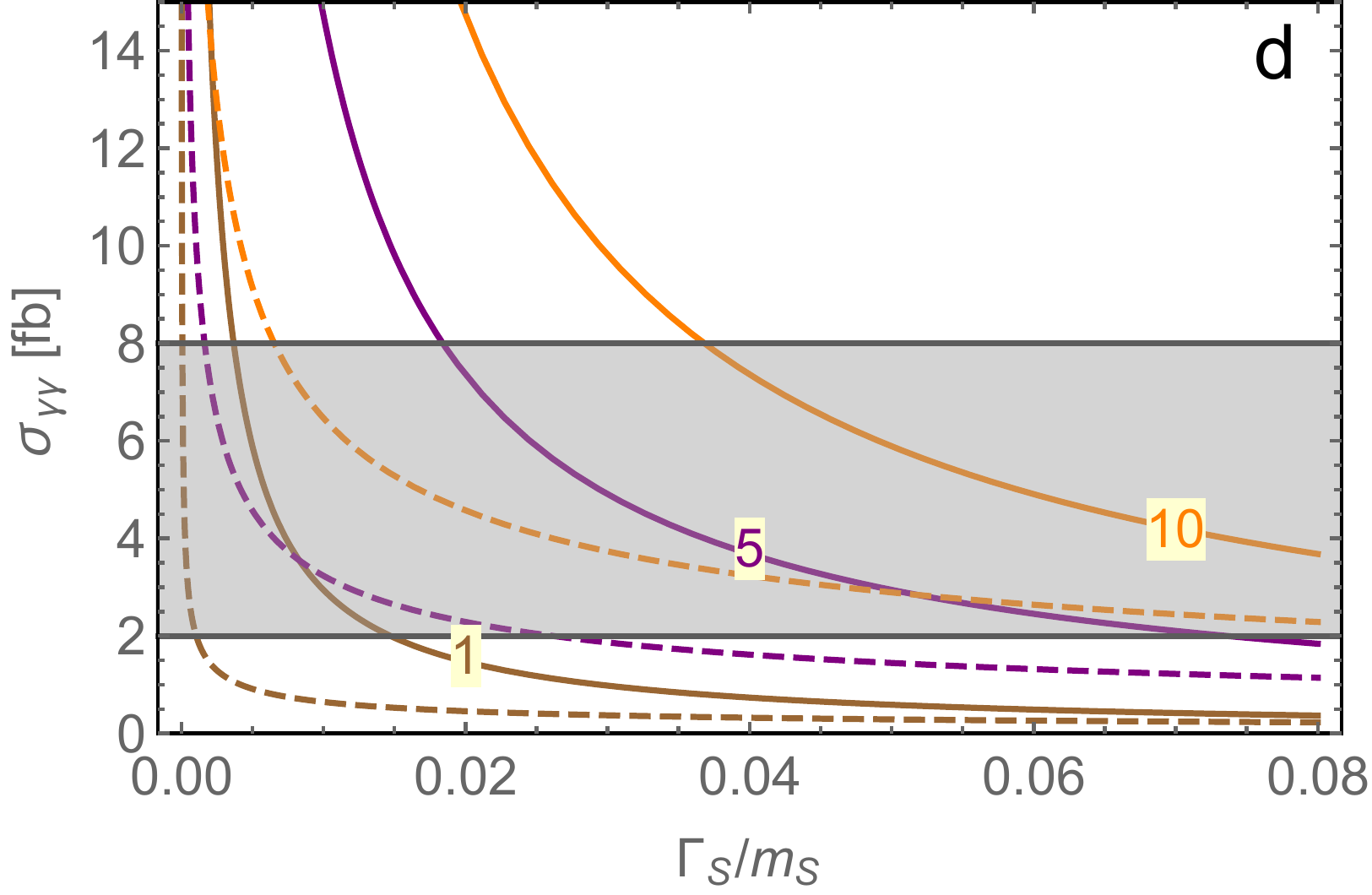}
\caption{Flavor and dijet constraints on a tree-level 2HDM model for the 750~GeV resonance, assuming $m_H= m_A$ and summing the contributions of both. The x-axis and y-axis correspond to the total width $\Gamma_S/M_S$ and diphoton production cross section $\sigma(pp\to S\to\gamma\gamma)$, respectively. Grey horizontal band marks $\sigma(pp\to S\to\gamma\gamma)$ required to address the diphoton excess. The width hint at ATLAS corresponds to $\Gamma_S/M_S=0.06$. The constraints are expressed as contours of constant $\frac{\Gamma(S\to\gamma\gamma)/M_S}{\left(\alpha/4\pi\right)^2}=1,5,10$, with solid (dashed) lines representing flavor (dijet) limits. Going above a given contour means that larger value of $\frac{\Gamma(S\to\gamma\gamma)/M_S}{\left(\alpha/4\pi\right)^2}$ would be needed in order to accommodate the corresponding point in the $\Gamma_S/M_S\;-\;\sigma_{\gamma\gamma}$ plane, due to the corresponding experimental constraint. Left: diagonal coupling to $uu^c$, represented by $(Hq)Yu^c$ in Eq.~(\ref{eq:treeL}) with $Y=(y,0,0)$. Right: diagonal coupling to $dd^c$, represented by replacing Eq.~(\ref{eq:treeL}) with $H^\dag qYd^c$ with $Y=(y,0,0)$.}
\label{fig:treeuu}
\end{figure}

In the next two subsections we examine flavor and collider constraints, finding that, because of these constraints on the production vertex, large partial width to diphotons $\frac{\Gamma(S\to\gamma\gamma)/M_S}{\left(\alpha/4\pi\right)^2}>1$ would be needed for the tree-level coupling 2HDM to explain the 750~GeV resonance. Before going into the detailed analyses we summarise the results in Fig.~\ref{fig:treeuu}, that we now explain.

Fig.~\ref{fig:treeuu} shows contours of constant $\frac{\Gamma(S\to\gamma\gamma)/M_S}{\left(\alpha/4\pi\right)^2}$ in the diphoton cross section (y-axis)-signal width (x-axis) plane. The  value of $\frac{\Gamma(S\to\gamma\gamma)/M_S}{\left(\alpha/4\pi\right)^2}$ indicated by the contours is the {\it lower limit} that is needed to match the corresponding values of cross section and total width.

The left (right) panels in Fig.~\ref{fig:treeuu} show the results when considering diagonal Yukawa couplings of $H$ to the first generation up (down) type quarks\footnote{We comment in the next subsections about the implications of coupling $H$ to both first and second generation quarks in a flavor U(2)-symmetric way; as we show, this possibility, while it ameliorates the flavor constraints associated with breaking the approximate light flavor U(2) symmetry of the SM, makes the dijet constraints correspondingly stronger and so does not affect our conclusions.}. The grey band marks the cross section required to explain the ATLAS and CMS excess. Solid lines show the constraint arising from flavor [Eq.~(\ref{eq:sigtreeu}) and discussion around it], while dashed lines show the constraint due to dijet searches [Eq.~(\ref{dijetLHC}) and discussion around it].

To substantiate the theory difficulty in obtaining such large diphoton partial width, $\frac{\Gamma(S\to\gamma\gamma)/M_S}{\left(\alpha/4\pi\right)^2}>1$, in App.~\ref{app:2hdm} we explore the associated auxiliary model-building gymnastics in two concrete examples. In App.~\ref{aapp:1} we show that the charged Higgs loop of the $H^\pm$ state contained in $H$ can make only a negligible contribution to the diphoton partial width of $H^0$ and $A^0$, compared to the required $\frac{\Gamma(S\to\gamma\gamma)/M_S}{\left(\alpha/4\pi\right)^2}>1$. In App.~\ref{aapp:2} we add vector-like fermions to boost the diphoton decay, and show that extreme limits of the parameters would be needed to achieve the required effect.

\subsubsection{Flavor constraints}
Any model with extra Higgs doublets, and in particular the model in Eq.~(\ref{eq:treeL}), has flavor changing couplings for either the neutral scalar or the charged scalar or both. Consequently, our model is constrained by measurements of $K^0-\overline{K}{}^0$ and $D^0-\overline{D}{}^0$ mixing \cite{Blum:2011fa}, where box diagrams containing the charged and/or neutral states in $H$ lead to contributions that we summarize by:
\begin{itemize}
\item $D^0-\overline{D}{}^0$ mixing:
\begin{equation}\label{eq:DD}
\frac{1}{32\pi^2}\left(\frac{{\rm TeV}}{M_H}\right)^2\sum_{ij}
Y_{1i}Y_{2i}^*Y_{1j}Y_{2j}^*<10^{-5}.
\end{equation}
\item $K^0-\overline{K}{}^0$ mixing:
\begin{equation}\label{eq:KK}
\frac{1}{32\pi^2}\left(\frac{{\rm TeV}}{M_H}\right)^2\sum_{ij}
(V^\dagger Y)_{1i}(V^\dagger Y)_{2i}^*(V^\dagger Y)_{1j}(V^\dagger Y)_{2j}^*<2\times10^{-5}.
\end{equation}
\end{itemize}
We took the zero mass limit for internal quarks in the loop function.

There are two limiting cases where the contributions to flavor changing neutral current processes appear in only one of the two quark sectors. (In a more generic case, the constraints are similar, typically stronger.) First, consider $Y={\rm diag}(y,0,0)$. In this case, the charged scalar couplings to $d_{Li}^\dagger u_{Rj}$ are given by $(V^\dagger Y)_{ij}=y V^*_{1j}\delta_{i1}$. Box diagrams, involving the exchange of $H^\pm$ contribute to $K^0-\overline{K}{}^0$ mixing. We obtain:
\begin{equation}
\label{eq:flav}y<0.5\times[M_{H}/750\ {\rm GeV}]^{1/2}.
\end{equation}
Second, consider $V^\dagger Y={\rm diag}(y,0,0)$. In this case, the neutral scalar couplings to $u_{Li}^\dagger u_{Rj}$ are given by $Y_{ij}=y V_{i1}\delta_{j1}$. Box diagrams, involving the exchange of the neutral scalars contribute to $D^0-\overline{D}{}^0$ mixing, and give us again a   limit quantitatively similar to Eq.~(\ref{eq:flav}).

Using Eq.~(\ref{eq:flav}) in Eq.~(\ref{eq:1}) we have
\be\label{eq:siguuH}
\sigma(pp\to H^0)\lesssim 60~{\rm pb}
\ee
with a similar bound for $A^0$.
Using now Eq.~(\ref{eq:siguuH}) in Eq.~(\ref{eq:sigtree}), summing the contributions of $H^0$ and $A^0$ and denoting the sum collectively by $\sigma(pp\to S\to\gamma\gamma)$, etc., we find
\be\label{eq:sigtreeu}
\left[\frac{\Gamma(S\to\gamma\gamma)/M_S}{\left(\alpha/4\pi\right)^2}\right]\gtrsim6~\left(\frac{K}{2}\right)^{-1}\left[\frac{\sigma(pp\to S\to\gamma\gamma)}{5~{\rm fb}}\right]\times\left[\frac{\Gamma_S/M_S}{0.06}\right].
\ee
Eq.~(\ref{eq:sigtreeu}) is depicted in the left panel of Fig.~\ref{fig:treeuu} as solid contours of constant $\Gamma(S\to\gamma\gamma)/M_S$ in the $\Gamma_S/M_S-\sigma(pp\to S\to\gamma\gamma)$ plane.

We find results comparable to Eq.~(\ref{eq:sigtreeu}) when considering a dominant coupling of $H^0$ to down quarks, obtained by replacing $(Hq)Yu^c$ in the first line of Eq.~(\ref{eq:treeL}) with $H^\dag qYd^c$. The main difference is that the constraint analogous to Eq.~(\ref{eq:sigtreeu}) becomes somewhat tighter, because of the smaller PDF for $d\bar d$ compared to $u\bar u$ at the LHC. This effect for the $H^0dd^c$ scenario can be captured by replacing $K\to0.6K$ in Eq.~(\ref{eq:sigtreeu}). The result is shown in the right panel of Fig.~\ref{fig:treeuu}.
We note that constraints similar to Eq.~(\ref{eq:flav}) are obtained when coupling $H^0$ to any one particular case of the eight pairs of first two generations quarks, such as $c u^c$ or $d s^c$. However, going to, e.g., $uc^c$ or $ds^c$ for the initial state introduces a significant PDF suppression in the production vertex, making the corresponding version of Eq.~(\ref{eq:sigtreeu}) much more constraining.

Up to here we focused on dominant coupling of $H^0$ to one particular pair of quarks. The flavor constraint thus derived manifests the breaking that such a coupling introduces to the approximate U(2) flavor symmetry of the first two generations in the SM. This suggests a way out, by assuming $Y= {\rm diag}(y,y,0)$ or $V^\dag Y={\rm diag}(y,y,0)$ in Eq.~(\ref{eq:treeL}) (or similarly for the $H^\dag qYd^c$ case)\footnote{We are grateful to Gilad Perez for a discussion of this point.}. Adopting this form of $Y$, both Eqs.~(\ref{eq:DD}-\ref{eq:KK}) vanish up to CKM and GIM suppression factors that make flavor violation unobvservably small. We conclude that it is possible in principle to avoid the flavor constraint on $y$ by imposing flavor U(2) involving the first two generation quarks. The cost of the U(2) limit, however, is a factor $\sim2$ enhancement in the dijet decay width of $H^0$. This is shown in the next section to lead to strong constraints on its own.

Finally, we emphasize that in Eq.~(\ref{eq:sigtreeu}) we took the contributions of $H^0$ and $A^0$ to the $S\to\gamma\gamma$ signal to be equal and, moreover, assumed that $M_{H^0}$ and $M_{A^0}$ are equal up to a splitting smaller than or comparable to the $m_{\gamma\gamma}$ resolution of the CMS and ATLAS analysis algorithms. In slight more generality, there is no reason a-priori for the Yukawa coupling $Y$ in Eq.~(\ref{eq:treeL}) to be real, in which case the contributions of $H^0$ and $A^0$ would be different and include interference effects. Moreover, a mass splitting $M_{H^0}-M_{A^0}$ comparable to the $m_{\gamma\gamma}$ resolution of ATLAS and CMS would lead to the two states contributing to separate, and individually less significant, resonances. Generically, we expect that this will result in a stronger constraint than Eq.~(\ref{eq:sigtreeu}), by about a factor of 2.

Along the same line, we note that nearly-degenerate scalar and pseudo-scalar components of a neutral complex scalar field, despite having individually narrow widths, could perhaps mimic a wider bump as hinted to by the ATLAS analysis. This is an interesting way to potentially get away from the tight spot associated with an experimentally deduced large width, though the precise implementation would require knowledge of the effects in the ATLAS and CMS bump hunt algorithms. While we do not pursue this possibility further, we note that splitting the two states by more than about 10~GeV should allow ATLAS and CMS, once better statistics become available, to resolve the two adjacent bumps with their current electromagnetic energy resolution.

\subsubsection{Dijet constraints}

The tree-level couplings of the 2HDM are further constrained by the LHC dijet resonance search at 8~TeV, which can be directly translated to the production cross-section at 13~TeV. We use MSTW2008 pdf set at NNLO~\cite{Martin:2009iq} to find
\be
\frac{\mathcal{L}_{13}^{pp}}{\mathcal{L}_{8}^{pp}}
\simeq\begin{cases}
2.5 & u\bar u{\rm ~coupling}, \\
2.7 & d\bar d{\rm ~coupling}. \end{cases}
\ee
The 8~TeV dijet results are reported as $95\%$ C.L. limit on $\sigma_{8}(pp\to S\to jj)\times A$, where $A$ is an acceptance factor reflecting the different cuts used by ATLAS and CMS. Estimating the acceptance as $A\simeq0.5$ using MG5, we approximate the dijet bound, at 750~GeV, as $\sigma_{8}(pp\to S\to jj)\leq2$~pb~\cite{Aad:2014aqa,CMS-PAS-EXO-14-005}. Summing again the contributions of $H^0$ and $A^0$ we thus find
\be\label{dijetLHC}
\left[\frac{\Gamma(S\to\gamma\gamma)/M_S}{(\alpha/4\pi)^2}\right]\gtrsim14\left(\frac{\mathcal{L}_{13}/\mathcal{L}_{8}}{2.5}\right)^{-1/2}\left(\frac{K}{2}\right)^{-1/2} \left[\frac{\sigma_{13}(pp\to S\to\gamma\gamma)}{5{\rm~fb}}\right]\times\left[\frac{\Gamma_S/M_S}{0.06}\right]^{1/2}.
\ee
The resulting constraint is depicted in dashed lines in Fig.~\ref{fig:treeuu}.

A comparable bound is provided by the TeVatron dijet resonance search at 1.96~TeV. When coupled to first generation quarks, the production at the TeVatron exhibits a valence-PDF enhancement compared to the LHC, that partially compensates for the smaller CoM energy. The luminusity ratios are in this case
\be
\frac{\mathcal{L}_{13}^{pp}}{\mathcal{L}_{1.96}^{p\bar p}}
\simeq\begin{cases}
35 & u\bar u{\rm ~coupling}, \\
250 & d\bar d{\rm ~coupling}. \end{cases}
\ee
The Tevatron dijet results are reported as $95\%$ C.L. limit on $\sigma_{1.96}(p\bar p\to S\to jj)\times A$, for central $(|y|<1)$ jets. We again estimate the acceptance using MG5 and find $A\simeq0.6$. The Tevatron dijet bound, at 750~GeV, then reads $\sigma_{1.96}(p\bar p\to S\to jj)\leq1$~pb~\cite{Aaltonen:2008dn}, resulting in
\be\label{tev}
\left[\frac{\Gamma(S\to\gamma\gamma)/M_S}{(\alpha/4\pi)^2}\right]\gtrsim6\left(\frac{\mathcal{L}_{13}^{pp}/\mathcal{L}_{1.96}^{p\bar p}}{35}\right)^{-1/2}\left(\frac{K}{2}\right)^{-1/2}\left[\frac{\sigma_{13}(pp\to S\to\gamma\gamma)}{5{\rm~fb}}\right]\times\left[\frac{\Gamma_S/M_S}{0.06}\right]^{1/2}.
\ee

Finally, referring back to the U(2) flavor symmetry mentioned in the previous section as means to avoid flavor constraints, we note that adopting the U(2) limit would amount, roughly, to replacing $K\to 0.5K$ in Eqs.~(\ref{dijetLHC}) and~(\ref{tev}), due to the larger dijet decay width of the scalar.

\section{Conclusions}\label{sec:conc}
The ATLAS and CMS experiments have reported a hint for a resonant excess of diphoton events around 750 GeV. We study the possibility that this excess, if not a statistical fluctuation, is generated by a spin-0 resonance $S$. In this case, the data imply a cross section $\sigma(pp\to S\to\gamma\gamma)$ of a few fb.

We argued, in a {\it model-independent way}, that new charged particles, beyond the Standard Model (SM) ones, are necessary in order to give a large enough $S$ decay rate into two photons. Concretely, we use data from Run-I of the LHC, showing that neither of the two heavy charged SM particles -- the top quark and the $W$-boson -- can generate $\sigma(pp\to S\to\gamma\gamma)$ that is large enough. (The light charged SM particles give negligible contributions.)

Intriguingly, the ATLAS data show preference for a large width, of order $\Gamma_S\sim 45\ {\rm GeV}\sim0.06 M_S$. Our focus in this work was on the question of whether the combination of the three observables, $M_S$, $\sigma(pp\to S\to\gamma\gamma)$ and, in particular, large $\Gamma_S$, can be accounted for in reasonable theoretical scenarios.

We first studied the scenario where the dominant production mechanism of $S$ is gluon fusion (ggF). We showed that if (i) the number of new particles is of order one, (ii) their electromagnetic charge is of order one, and (iii) their coupling to $S$ is of order one, then the rate of diphoton events would be 2-3 orders of magnitude too small. Thus, only rather exotic models, employing large multiplicity of new states, or large charges, or large couplings, or some combination of these unusual features, can explain the data.

We then considered tree level production, $q\bar q\to S$. We showed that constraints from flavor changing neutral current processes and from collider dijet analyses put an upper bound on the production cross section. Again, the new charged particles that lead to the $S\to\gamma\gamma$ decay should be rather exotic, as they need to generate $\Gamma(S\to\gamma\gamma)/M_S\gtrsim 5(\alpha/4\pi)^2$ if one aims to accommodate diphoton production rate and bump width in the ballpark of the best fit values. We demonstrated in specific models the difficulty in achieving such a high rate.

Our final conclusion is that a large total width is very difficult to accommodate in perturbative extensions of the Standard Model. It remains to be seen whether the diphoton excess will survive larger statistics and, if it does, whether the large width will still be implied by the measurements.\\

{\it Note added in proof.} The presentation of a 750~GeV excess by the ATLAS and CMS collaborations~\cite{ATLAS-CONF-2015-081,CMS-PAS-EXO-15-004} triggered many analyses by the particle physics community.
A few selected published examples, with results corroborating and partially overlapping to ours, include~\cite{Buttazzo:2015txu,Low:2015qep,Franceschini:2015kwy,Ellis:2015oso,Falkowski:2015swt}; many more analyses can be found in the references to~\cite{ATLAS-CONF-2015-081,CMS-PAS-EXO-15-004}. To the best of our knowledge, none of the works includes our flavor analysis, and none analyzed the implications of a large total $S$ width in the detailed and general way we attempted here. 

\acknowledgments
We thank Gilad Perez for useful discussions on v2 of this manuscript. YN is the Amos de-Shalit chair of theoretical physics. This research is supported by the I-CORE program of the Planning and Budgeting Committee and the Israel Science Foundation (grant number 1937/12). YN is supported by a grant from the United States-Israel Binational Science Foundation (BSF), Jerusalem, Israel.

\begin{appendix}
\section{Fermion and vector boson loop amplitudes in $S\to\gamma\gamma$}\label{app:a}
The fermionic loop amplitude in the effective coupling of $S$ to photons is given by (see e.g.~\cite{Djouadi:1993ji,Kniehl:1995tn})
\be\mathcal{A}_\psi(\tau)&=&\frac{\xi+2}{2\tau^2}\left(\xi\tau+(\tau-\xi)f(\tau)\right),\;\;\;\;\tau\equiv \left(\frac{M_S}{2m_\psi}\right)^2,\\
f(\tau)&=&\left\{\begin{array}{cc}\arcsin^2\sqrt{\tau},&\tau\leq1\\-\frac{1}{4}\left(\log\frac{1+\sqrt{1-\tau^{-1}}}{1-\sqrt{1-\tau^{-1}}}-i\pi\right)^2,&\tau>1\end{array}\right\},\ee
with $\xi=1(0)$ for a scalar (pseudo-scalar).
The vector boson amplitude for scalar $S$ is given by
\be\mathcal{A}_V(\tau)&=&-\frac{1}{\tau^2}\left(3\tau+2\tau^2+3(2\tau-1)f(\tau)\right),\;\;\;\;\tau\equiv \left(\frac{M_S}{2m_V}\right)^2,\ee
with the same function $f(\tau)$.

\section{2HDM diphoton decays}\label{app:2hdm}
In Sec.~\ref{ssec:treewide} we showed that tree-level production of $S$, that could be achieved in principle by associating $S$ with the scalar and/or pseudo-scalar components of the neutral part of a heavy Higgs doublet, would require the model to satisfy the constraint $\frac{\Gamma(S\to\gamma\gamma)/M_S}{\left(\alpha/4\pi\right)^2}>1$ in order to explain the ATLAS and CMS 750~GeV diphoton excess. In the next two subsections we illustrate the parametric difficulty in satisfying this constraint.

\subsection{Diphoton decay via Higgs mixing and charged Higgs loop}\label{aapp:1}
The first point we clarify is that the loop diagram involving the $H^\pm$ state cannot (by far) accommodate $\frac{\Gamma(S\to\gamma\gamma)/M_S}{\left(\alpha/4\pi\right)^2}>1$.

For the charged Higgs loop to contribute to $H\to\gamma\gamma$, and given that we must not allow a large VEV for $H$, we note that mixing with the SM state $h$ is needed. We can obtain this by turning on
\be\mathcal{L}&=&-\lambda_7H^\dag h|H|^2+{\rm h.c.}\no\\
&=&-\left(\Re\lambda_7H^0-\Im\lambda_7A^0\right)\left(h+v\right)\left(H^+H^-+\frac{H^{02}+A^{02}}{2}\right),\ee
where $h(1,2)_{\frac{1}{2}}=\left(0,\frac{h+v}{\sqrt{2}}\right)^T$ is the SM Higgs in Unitary gauge (with some abuse of notation we denote by $h$ both the full SM doublet and its fluctuating real scalar component).

Neglecting the VEV of $H$, the field-dependent tree level charged Higgs mass is
\be M_{H^\pm}^2&=&M_H^2+\left(\Re\lambda_7H^0-\Im\lambda_7A^0\right)(h+v).\ee
We can use the Higgs low energy theorem (LET, see e.g.~\cite{Kniehl:1995tn}) to compute the leading-log diphoton amplitude,
\be\mathcal{L}&=&-\frac{\alpha}{8\pi v}\left(\mathcal{A}_{s,H^\pm}H^0FF+\frac{3}{4}\mathcal{A}_{p,H^\pm}A^0F\tilde F\right),\ee
with
\be \mathcal{A}_{s,H^\pm}&=&v\,b_{H}\frac{\partial \log M_{H^\pm}}{\partial H^0}=\frac{\Re\lambda_7}{6}\frac{v^2}{M^2_{H^\pm}},\no\\
\mathcal{A}_{p,H^\pm}&=&v\,b_{H}\frac{\partial \log M_{H^\pm}}{\partial A^0}=-\frac{\Im\lambda_7}{6}\frac{v^2}{M^2_{H^\pm}},
\ee
where $b_{H}=\frac{1}{3}$. This gives the diphoton width
\be\Gamma(H^0\to\gamma\gamma)&\approx&\left(\frac{\alpha}{4\pi}\right)^2\frac{\Re\lambda_7^2}{576\pi}\left(\frac{M_H}{M_{H^\pm}}\right)^4\left(\frac{v}{M_H}\right)^2M_H\ll\left(\frac{\alpha}{4\pi}\right)^2M_H,\ee
as expected, with a similar result for $A^0$.

\subsection{Diphoton decay via vector-like fermions}\label{aapp:2}
Let us introduce vector-like leptons $\psi(1,2)_{-\frac{1}{2}}$, $\psi^c(1,2)_{+\frac{1}{2}}$, $\chi(1,1)_{+1}$, $\chi^c(1,1)_{-1}$, with
\be\label{eq:10}\mathcal{L}&=&-M_\psi\psi\psi^c-M_\chi\chi\chi^c-\left\{y_HH^\dag \psi\chi+y_H^c(H\psi^c)\chi^c+y_hh^\dag \psi\chi+y_h^c(h\psi^c)\chi^c+{\rm h.c.}\right\}.\ee
For simplicity we take all masses and couplings to be real. We will use the Higgs LET to estimate the induced diphoton decay width of $H$. With real couplings, the LET yields an effective $H^0FF$ vertex, but no analogous $A^0F\tilde F$ vertex. Instead, the $A^0F\tilde F$ vertex can be obtained from the ABJ anomaly as we utilise below.
For simplicity we assume $M_\psi M_\chi+\frac{y_hy_h^cv^2}{2}>0$, setting $M_\psi M_\chi+\frac{y_hy_h^cv^2}{2}=M_1M_2$, where $M_1$ and $M_2$ are the positive charged Dirac fermion mass eigenvalues. Allowing for $N$ identical copies of~(\ref{eq:10}) with
vector-like fermion masses larger than $M_H/2$, we then have
\be\mathcal{L}&=&\frac{\alpha}{8\pi v}N\mathcal{A}_sH^0FF,\ee
where
\be \mathcal{A}_s&=&v\,b_{f}\frac{\partial \log {\rm Det}M_{f}}{\partial H^0}=\frac{2(y_hy^c_H+y_h^cy_H)\,v^2}{3M_1M_2},
\ee
with $b_{f}=\frac{4}{3}$. This gives the diphoton partial width:
\be\Gamma(H^0\to\gamma\gamma)&=&\frac{\alpha^2M_H^3}{256\pi^3v^2}N^2\left|\mathcal{A}_{s}\right|^2
=\left(\frac{\alpha}{4\pi}\right)^2N^2\,\frac{|y_hy^c_H+y_h^cy_H|^2}{36\pi}\left(\frac{M_H^2}{M_1M_2}\right)^2\left(\frac{v}{M_H}\right)^2M_H.\ee
For the pseudo-scalar we have~\cite{Kniehl:1995tn}
\be\mathcal{L}&=&\frac{3}{4}\frac{\alpha}{8\pi v}N\mathcal{A}_pA^0F\tilde F,\ee
with $\mathcal{A}_p\approx\mathcal{A}_s$,
giving
\be\Gamma(A^0\to\gamma\gamma)&=&\frac{9\alpha^2M_H^3}{512\pi^3v^2}N^2\left|\mathcal{A}_{p}\right|^2
=\left(\frac{\alpha}{4\pi}\right)^2N^2\,\frac{|y_hy^c_H+y_h^cy_H|^2}{16\pi}\left(\frac{M_H^2}{M_1M_2}\right)^2\left(\frac{v}{M_H}\right)^2M_H.\ee

Once again tuning $M_{H^0}\sim M_{A^0}$ to within the $m_{\gamma\gamma}$ resolution of the ATLAS and CMS analyses, the effective decay width [assuming doubled total production cross section as in Eq.~(\ref{eq:sigtreeu})] is
\be \left[\frac{\Gamma(S\to\gamma\gamma)/M_S}{\left(\alpha/4\pi\right)^2}\right]_{eff}&\approx&\frac{13N^2}{88\pi}|y_hy^c_H+y_h^cy_H|^2\left(\frac{M_H^2}{M_1M_2}\right)^2\left(\frac{v}{M_H}\right)^2\no\\
&\lesssim&0.1y^4N^2,\ee
collectively denoting $y^4\equiv|y_hy^c_H+y_h^cy_H|^2$. In the last line, to obtain a conservative estimate, we set $M_H^2/(M_1M_2)=4$, letting at least one of the charged vector-like fermions be lighter than 350~GeV.

With three identical copies, $N=3$, Yukawa couplings $y\sim\mathcal{O}(1)$, and tuned parameters so that $M_2\approx M_1$ despite large mixing, we can just barely obtain the required diphoton decay width.

\end{appendix}


\bibliography{ref}


\end{document}